\documentclass[journal]{IEEEtran}
\usepackage[varg]{txfonts}
\usepackage{graphicx}
\usepackage{fixltx2e}
\usepackage{bm}
\usepackage{textcomp}
\usepackage{url}
\usepackage[english]{babel}
\usepackage[tracking,spacing,kerning,letterspace=0]{microtype}
\usepackage{wasysym}
\usepackage{balance}
\makeatletter
\renewcommand{\fnum@figure}{\figurename~\oldstylenums{\thefigure}}

\renewcommand\@biblabel[1]{[\oldstylenums{#1}]}
\makeatother
\fussy
\usepackage{memhfixc}

\begin{document}
\title{Spherical \textsc{gem}s for parallax-free detectors}
\author{\IEEEauthorblockN{Serge Duarte Pinto\IEEEauthorrefmark{1}\IEEEauthorrefmark{2}\IEEEauthorrefmark{3}, Matteo Alfonsi\IEEEauthorrefmark{1}, Ian Brock\IEEEauthorrefmark{2}, Gabriele Croci\IEEEauthorrefmark{1}, Eric David\IEEEauthorrefmark{1},\\  Rui de Oliveira\IEEEauthorrefmark{1}, Leszek Ropelewski\IEEEauthorrefmark{1}, Miranda van Stenis\IEEEauthorrefmark{1}, Hans Taureg\IEEEauthorrefmark{1}, Marco Villa\IEEEauthorrefmark{1}\IEEEauthorrefmark{2}.}
% \thanks{Manuscript received, \oldstylenums{14} November \oldstylenums{2008}. This project was made possible by support from the \textsc{rd}\oldstylenums{51} collaboration for micropattern gas detectors.}
\thanks{\IEEEauthorblockA{\IEEEauthorrefmark{1}\textsc{Cern}, Geneva, Switzerland.}}
\thanks{\IEEEauthorblockA{\IEEEauthorrefmark{2}Physikalisches Institut der Universit\"at Bonn, Bonn, Germany.}}
\thanks{\IEEEauthorblockA{\IEEEauthorrefmark{3}Corresponding author: Serge.Duarte.Pinto@cern.ch}}}

\maketitle
\thispagestyle{empty}

\noindent\begin{abstract}
\ We developed a method to make \textsc{gem} foils with a spherical geometry.
Tests of this procedure and with the resulting spherical \textsc{gem}s are presented.
Together with a spherical drift electrode, a spherical conversion gap can be formed.
This would eliminate the \emph{parallax error} for detection of x-rays, neutrons or \textsc{uv} photons when a gaseous converter is used. 
This parallax error limits the spatial resolution at wide scattering angles.
The method is inexpensive and flexible towards possible changes in the design.

We show advanced plans to make a prototype of an entirely spherical triple-\textsc{gem} detector, including a spherical readout structure.
This detector will have a superior position resolution, also at wide angles, and a high rate capability.
A completely spherical gaseous detector has never been made before.
\end{abstract}

\section{Motivation}
\label{intro}
\begin{figure}
\includegraphics[width=\columnwidth]{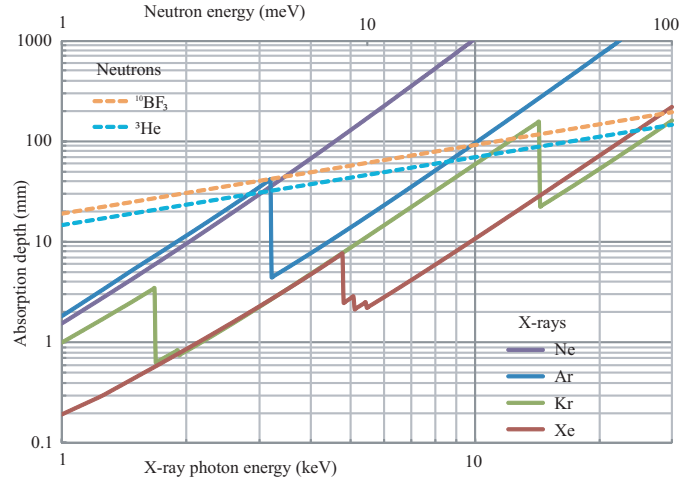}
\caption{Absorption length of various x-ray and thermal neutron conversion gases at atmospheric pressure, as a function of the energy of x-ray photons (lower horizontal scale) or neutrons (upper scale). Calculated from cross-section data~\cite{X-rayCross-sections, NeutronDataBooklet}.}
\label{Absorption_Gases}
\end{figure}

\begin{figure}[b]
\includegraphics[width=\columnwidth]{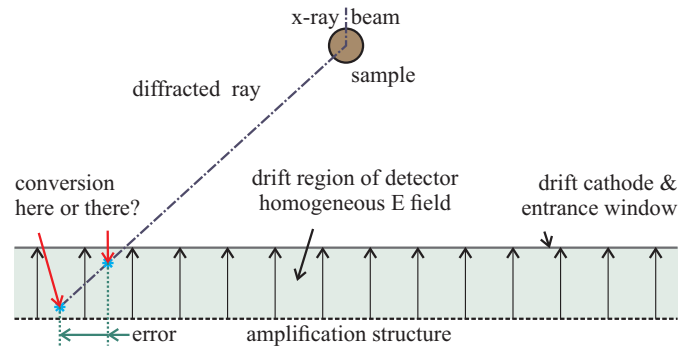}
\caption{The cause of a parallax error in a gas detector with a homogeneous drift field.}
\label{ParallaxError}
\end{figure}

\noindent Position sensitive radiation detectors with a spherical geometry can be attractive for various applications, for several reasons.
The most common reason in the case of gas detectors is to eliminate the \emph{parallax error} arising from the uncertainty of how deep radiation penetrates the sensitive volume before causing ionization.
Figure~\oldstylenums{\ref{Absorption_Gases}} shows that the depth of interaction of x-rays and thermal neutrons in a gas volume ranges from few to many millimeters, even when using the most efficient gases known for these purposes.
The situation is similar for gaseous \textsc{vuv} photoconverters such as tetrakis dimethylamine ethylene (\textsc{tmae}) or triethylamine (\textsc{tea})~\cite{CharpakParallax}.
If the electric field in the conversion region of a gas detector is not parallel to the direction of irradiation, an uncertain conversion depth will give rise to an error in position reconstruction, see Fig.~\oldstylenums{\ref{ParallaxError}}.

Methods that have been used to suppress parallax error include:
\begin{itemize}
\item Arranging small area flat detectors in such positions as to approximate a spherical shape \cite{LobsterPaper}.
\item Creating an almost spherical conversion region with foils and meshes, then transferring the charge to a planar wire chamber \cite{CharpakPaper, Comparison}.
\item Having a spherical cathode with an otherwise flat detector, while reducing the conversion depth by using an efficient conversion gas at a high pressure ($\sim 3$ bar) \cite{MoscowPaper, BrukerPaper}.
\item Imitating a spherical cathode by dividing a flat electrode into concentric circular segments and controlling the voltage applied to each sector with a resistive divider \cite{SegmentedCathodePaper}.
\item Using a pulsed radiation source or additional hardware to calculate the depth of interaction per event, then correct the position reconstruction~\cite{NeutronDiffraction}.
\end{itemize}
Each of these methods has its limitations.
In all cases the challenge of making a fully spherical detector, however desirable, is avoided.
We star\-ted an effort that should lead to the first fully spherical gas detector, based on spherical GEMs, cathode and readout board.
We developed a method that allows us to make a spherical \textsc{gem} from a flat standard \textsc{gem}~\cite{firstGEM}, apparently without affecting its properties significantly.

\begin{figure*}
\includegraphics[width=\textwidth]{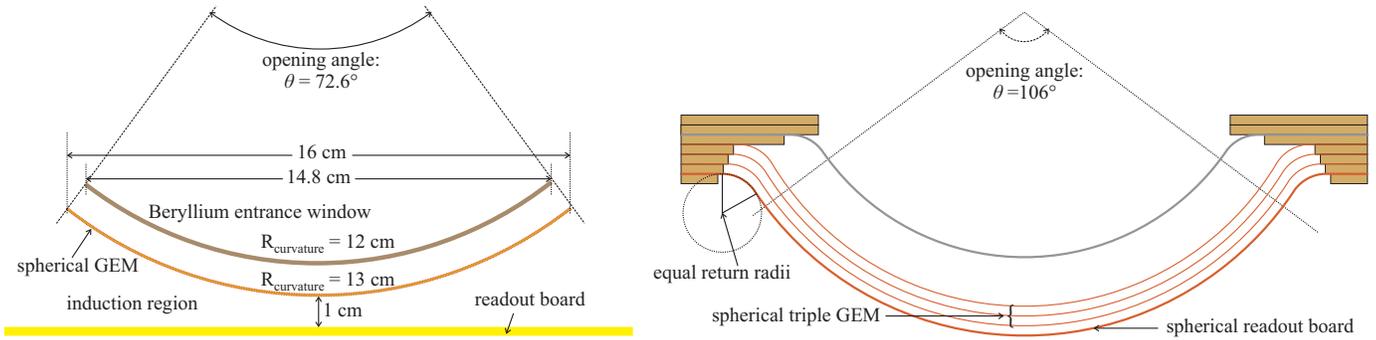}
\caption{Schematic composition of two prototypes based on spherical \textsc{gem}s. Left: single spherical \textsc{gem} with spherical drift electrode but flat readout structure. Right: an entirely spherical triple \textsc{gem} detector}
\label{BothDetectors}
\end{figure*}
First performance tests of single spherical \textsc{gem}s will be done in a setup with a spherical entrance window (which serves as drift electrode as well) and a flat readout structure, see the left side of Fig.~\oldstylenums{\ref{BothDetectors}}.
The electric field in the conversion region is truly radial.
The amplitude of signals from conversions in the non-radial induction region will be suppressed by the gain of the \textsc{gem}.

Designs are being prepared to make a prototype of a wide-angle spherical triple \textsc{gem}, see the right side of Fig.~\oldstylenums{\ref{BothDetectors}}.
\begin{figure*}
\begin{center}
\includegraphics[width=.9\textwidth]{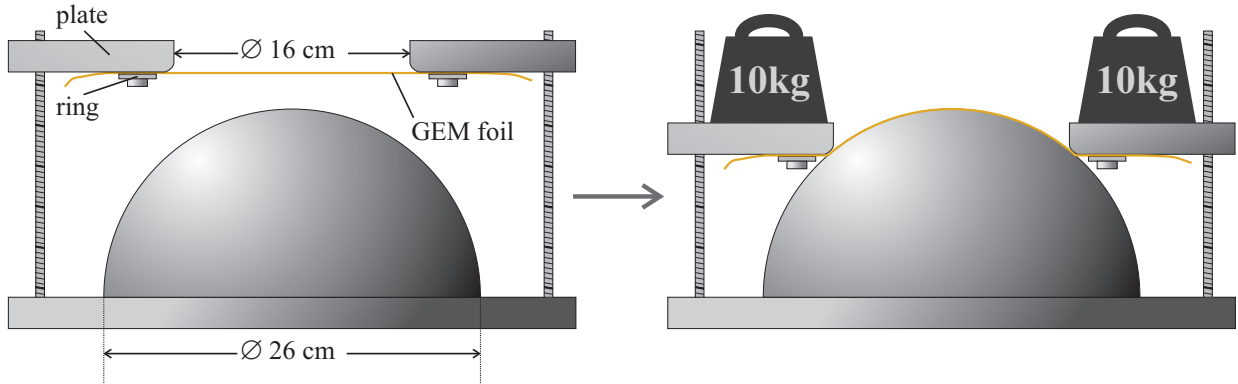}
\end{center}
\caption{The setup built to remold a flat \textsc{gem} into a spherical one. All part are stainless steel.}
\label{MoldingConstruction}
\end{figure*}
This will be the first entirely parallax-free gaseous detector.
Additional challenges of this design compared to the one on the left are the spherical readout structure and the narrow spacing between \textsc{gem}s, for which curved spacers are developed.

\section{Procedure \textit{\&} tooling}
\begin{figure}[h]
\center\includegraphics[width=.75\columnwidth]{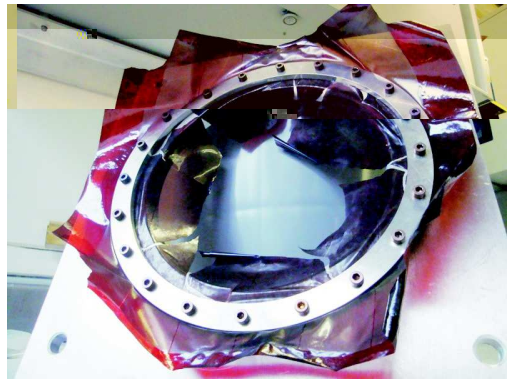}
\caption{The result of overheating a foil when trying to form it. The electrodes of this \textsc{gem} were removed beforehand to avoid their oxidation. During the forming procedure it was heated up to \oldstylenums{400}$^\circ$\,C.}
\label{TooHot}
\end{figure}

For the manufacturing of a spherical \textsc{gem} we start with a flat \textsc{gem} foil.
The shape of the electrodes is designed for the purpose; otherwise the foils used are no different from \textsc{gem}s used for other applications.
Thus, our base material is of proven reliability and we know its properties well.
Starting with this flat \textsc{gem} foil we use a method similar to thermoplastic heat forming; the foil is forced into a new shape by stretching it over a spherical mold, see Fig.~\oldstylenums{\ref{MoldingConstruction}}.
After a heat cycle it keeps this spherical shape.
Heat forming is routinely done with thermoplastic polymers, where above the so-called \emph{glass transition temperature} monomers can migrate freely, and polymerize again upon cooling down.
However the polyimide substrate of \textsc{gem}s is a thermoset polymer which has no well-defined glass transition temperature.
Strongly heating a polyimide leaves the polymer chains intact, but allows cross-links between chains to break or dislocate, thus relieving mechanical stresses.
Starting from a certain temperature the polyimide will start to degrade irreversibly; it becomes weak, brittle and dark-colored, see Fig.~\oldstylenums{~\ref{TooHot}}.
We found that \oldstylenums{350}$^\circ$\,C is the highest temperature we can apply to foils without damaging the quality of the polymer significantly.

Figure~\oldstylenums{\ref{MoldingConstruction}} shows the simple setup designed for forming \textsc{gem}s spherically.
It consists of a spherical mold, a ring-and-plate structure to hold the foil to be formed, and four rods along which the plate can slide down toward the mold.
Weights are applied to the plate to control the force that stretches the foil over the mold.
Note that, contrary to thermoplastic heat forming, the foil cannot be allowed to slip between the ring and plate where it is mounted.
The degree to which monomers migrate during the forming is much less than for thermoplastics; foils that accidentally slipped during forming were consistently wrinkled.
Also, the forming process for polyimide is apparently very slow: we found that the shortest heat cycle that gave satisfying results took \oldstylenums{24} hours.
\pagebreak

Partly due to this long heating time, the copper electrodes get fully oxidized in the process.
From such deep oxidation, electrodes cannot be recovered by etching.
The oxidation also causes some delamination of copper from the polyimide substrate.
Therefore the procedure should be carried out in an oxygen-free atmosphere.

\begin{figure}
\includegraphics[width=\columnwidth]{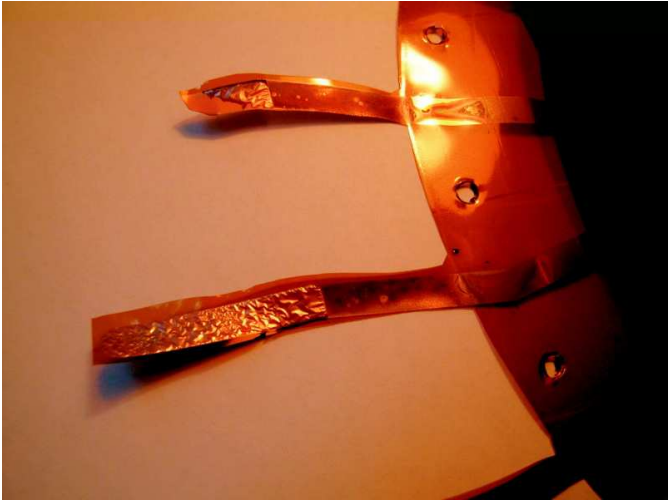}
\caption{Deposits on the electrode.}
\label{Deposits}
\end{figure}
Tests performed in a gas-tight enclosure with a constant flow of argon show that oxidation can indeed be largely avoided.
However, after forming the electrodes are covered with a thin film of an unknown substance, see Fig.~\oldstylenums{\ref{Deposits}}.
Samples taken from these deposits will be analyzed to identify the material and to understand the processes leading to its formation.
The very loose attachment to the electrodes suggests that outgassing of polyimide and/or copper plays a role.
These deposits can be removed afterwards, but as it involves mechanical brushing or rather strong spraying this is likely to affect the spherical shape and should better be avoided.

\begin{figure}
\includegraphics[width=\columnwidth]{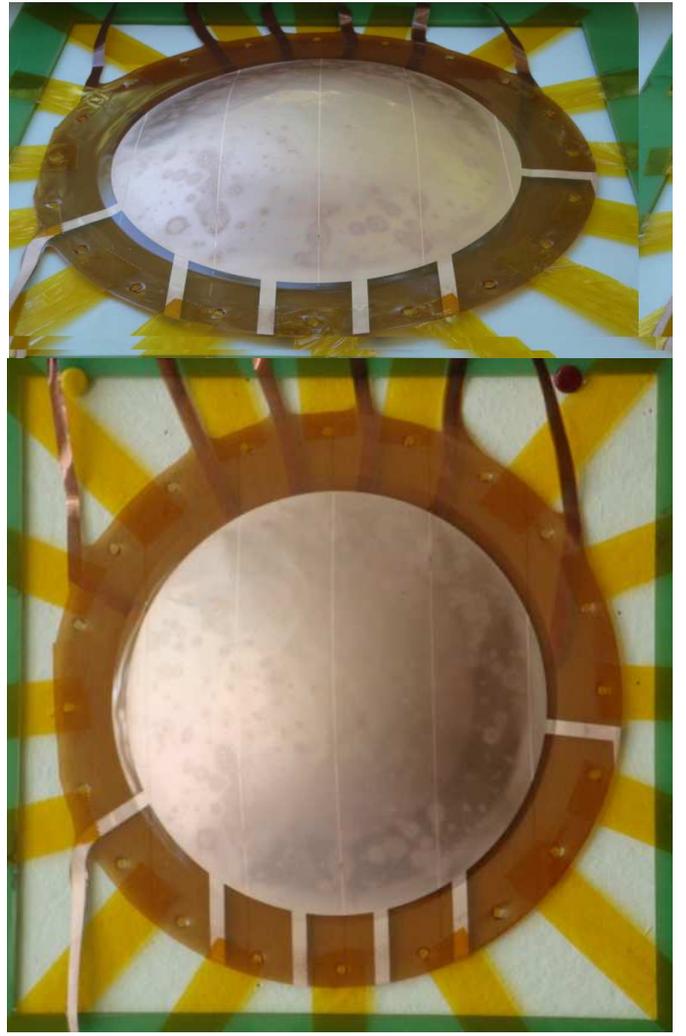}
\caption{Spherical \textsc{gem} made in a vacuum of $\sim$\,\oldstylenums{10} mbar.}
\label{UnderVacuum}
\end{figure}
To avoid formation of these deposits as much as possible the procedure should be carried out in a vacuum.
A test done in a moderate vacuum of $\sim$\,\oldstylenums{10} mbar gave encouraging results, see Fig.~\oldstylenums{\ref{UnderVacuum}}.
However, the foil still tends to attach to the mold, making it difficult to dismount the \textsc{gem} from the setup without damaging it.
A carbon spray coating of the mold is considered as an anti-adhesive layer to prevent this in the future.
In order to exclude outgassing of the polyimide to contribute to this attachment, foils are heated in a vacuum before they are mounted in the setup, until the outgassing process diminishes.
Outgassing studies on polyimides~\cite{outgassing} suggest that this point is reached after $\sim$\,\oldstylenums{10} hours, at a temperature of \oldstylenums{150}$^\circ$\,C.
This procedure need not be a great complication; contrary to the spherical forming this can be done with many foils at the same time.
In addition, \textsc{gem}s treated in this way have such benign outgassing behavior that they can be used in \emph{sealed} detectors, where the gas is not flowing~\cite{x-rayAstronomy}.
This opens the way to use more efficient gases such as xenon (see Fig.~\oldstylenums{\ref{Absorption_Gases}}), which would otherwise be prohibitively expensive.

\section{Properties of spherical \textsc{gem}s}
\textsc{Gem}s bent with the methods described above hold high voltages up to $\sim$\,\oldstylenums{650} V in air with a few nA leakage current, just like before the forming procedure.
When discharges occur, they are randomly distributed over the area of the foil.
This suggests that deformations due to the change in shape are spread homogeneously over the foil.
Observing holes at various locations on a spherical \textsc{gem} through a microscope confirms that, although they must have become wider, their shape has not become elliptical.

Assuming homogeneous stretching, one can estimate the change in relevant dimensions.
The active area of the foil before (flat) and after stretching (curved) is:
\begin{eqnarray}
A_\textrm{\scriptsize flat}  &=& \frac{\pi d^2}{4}=\pi r^2\sin^2\theta_{1/2}\\
A_\textrm{\scriptsize curved}&=& 2\pi \int_0^{\theta_{1/2}} r^2\sin\theta\textrm{d}\theta=2\pi r^2\left( 1-\cos\theta_{1/2} \right)
\end{eqnarray}
Where $r$ is the radius of curvature of the sphere, and $\theta_{1/2}$ is half the opening angle as indicated in Fig.~\oldstylenums{\ref{BothDetectors}}.
Then the surface stretching factor is:
\begin{equation}
\frac{A_\textrm{\scriptsize curved}}{A_\textrm{\scriptsize flat}}=2\frac{1-\cos\theta_{1/2}}{1-\cos^2\theta_{1/2}}
\end{equation}
It depends only on the opening angle.
If we substitute the numbers from Fig.~\oldstylenums{\ref{BothDetectors}}, we obtain \oldstylenums{10.7}\,\% and \oldstylenums{24.9}\,\% of increase in area for the left and right detectors respectively.

These stretching factors will have an effect on the aspect ratio of \textsc{gem} holes (defined as \emph{depth/width} of a hole), which in turn influences the amplifying behavior of the \textsc{gem}.
The change in width of a hole is proportional to the square root of the stretching factor, and the change in depth is inversely proportional to the stretching factor.
This results in aspect ratios reduced to \oldstylenums{86}\,\% and \oldstylenums{72}\,\% respectively, compared to their state before forming.
This will be compensated by changing the diameter and pitch of holes from the standard \oldstylenums{140}/\oldstylenums{70} microns to \oldstylenums{120}/\oldstylenums{60} and \oldstylenums{100}/\oldstylenums{50} microns respectively.

Another effect to be taken into account is the increase in capacitance of a foil due to stretching, as this will also increase the power generated in a discharge.
As the area of the foil increases with the stretching factor, the thickness of the dielectric decreases with the same factor.
Hence the capacitance increases with the square of the stretching factor: \oldstylenums{23}\,\% and \oldstylenums{56}\,\% respectively.
Capacitance measurements of the spherical \textsc{gem}s produced so far confirm these calculations.

\section{Work in progress}
\begin{figure}
\includegraphics[width=\columnwidth]{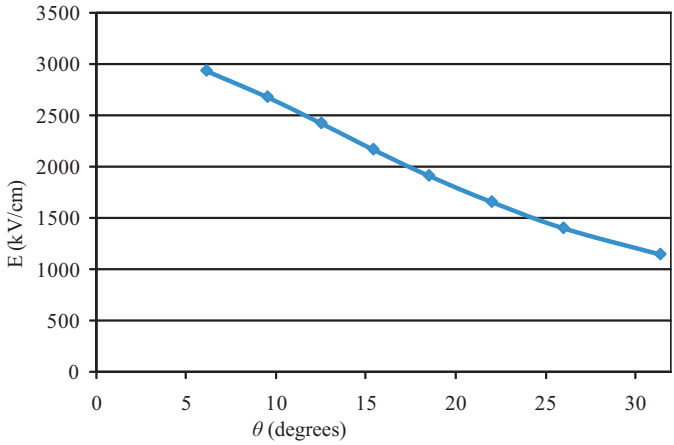}
\caption{Extraction field versus angle for the spherical \textsc{gem} with planar readout of Fig.~\oldstylenums{\ref{BothDetectors}}.}
\label{ExtractionField}
\end{figure}
\noindent At the time of writing the detectors shown in Fig.~\oldstylenums{\ref{BothDetectors}} are in the design phase.
Simulation studies are being made to understand the charge transfer between the spherical \textsc{gem} and the flat readout board (Fig.~\oldstylenums{\ref{BothDetectors}}, left).
For instance, Fig.~\oldstylenums{\ref{ExtractionField}} shows how the electrostatic field just below the spherical \textsc{gem} decreases at wider angle.
This is the extraction field for secondary electrons from the holes, which influences the effective gain of the \textsc{gem}.
The resulting $\theta$-dependence of the gain can be corrected for numerically.

\begin{figure}
\includegraphics[width=\columnwidth]{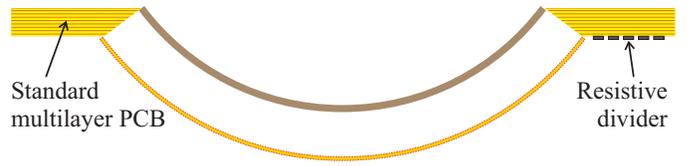}
\caption{The principle of a conical field cage made from a multilayer \textsc{pcb}.}
\label{FieldCage}
\end{figure}
\begin{figure}
\includegraphics[width=\columnwidth]{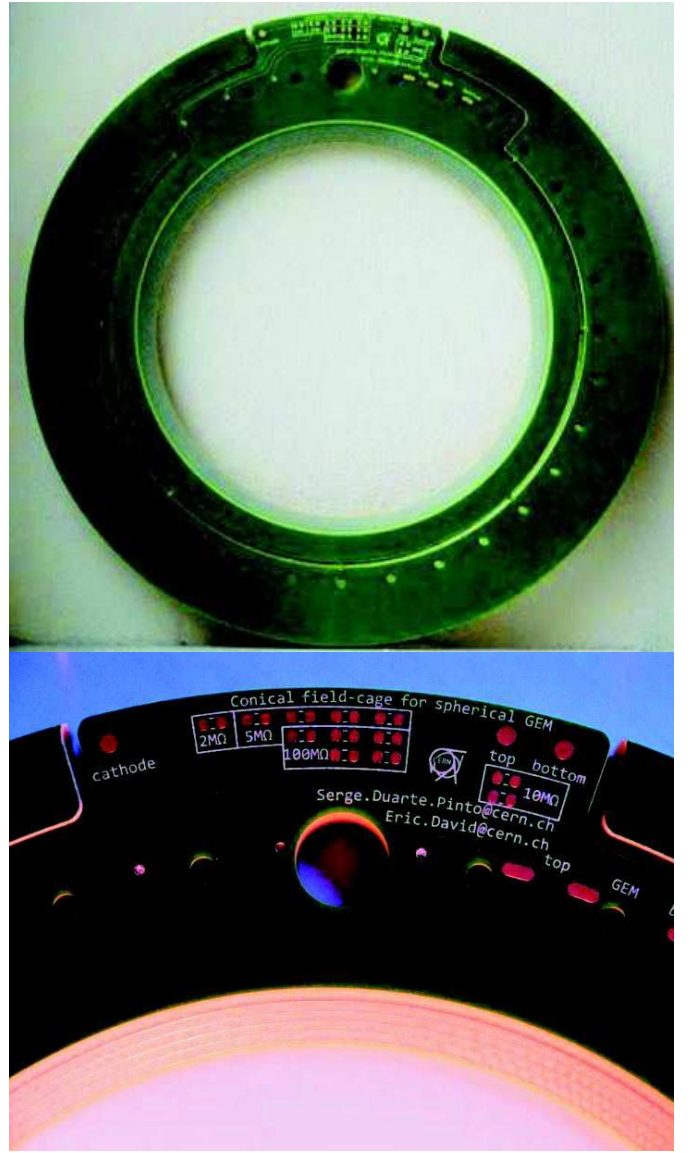}
\caption{The first conical field cage produced. Below a detail where the circuitry (without components) and the exposed annular electrodes are visible.}
\label{FirstFieldCage}
\end{figure}

The field quality in the drift region is more critical, as the elimination of the parallax error depends on it.
Therefore a spherical field cage is designed and manufactured to maintain a good radial field until the edge of the active area.
This field cage is a conical enclosure of the conversion region, and is made from a standard multilayer \textsc{pcb}, see Fig.~\oldstylenums{\ref{FirstFieldCage}}.
A resistive divider defines the voltages supplied to each layer.
It will also serve as a rigid mechanical fixture to which the \textsc{gem} can be glued, and as a high voltage distributor which supplies the \textsc{gem}.
\begin{figure}
\includegraphics[width=\columnwidth]{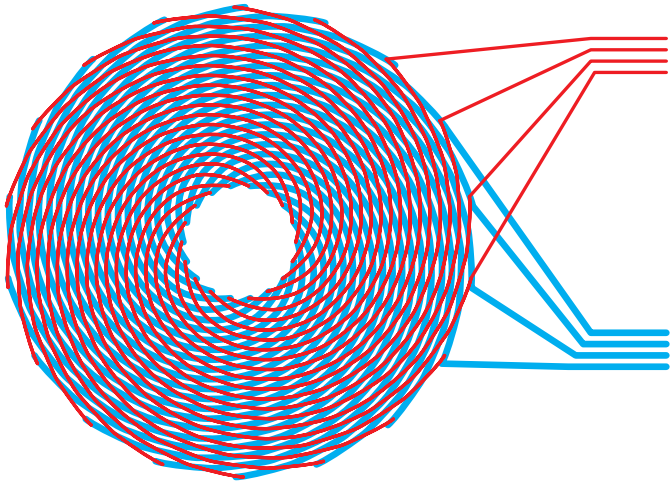}
\caption{Principle of spiral-pattern read\-out struc\-tu\-re for a spherical gas detector.}
\label{spirals}
\end{figure}
\begin{figure*}[t]
\includegraphics[width=\textwidth]{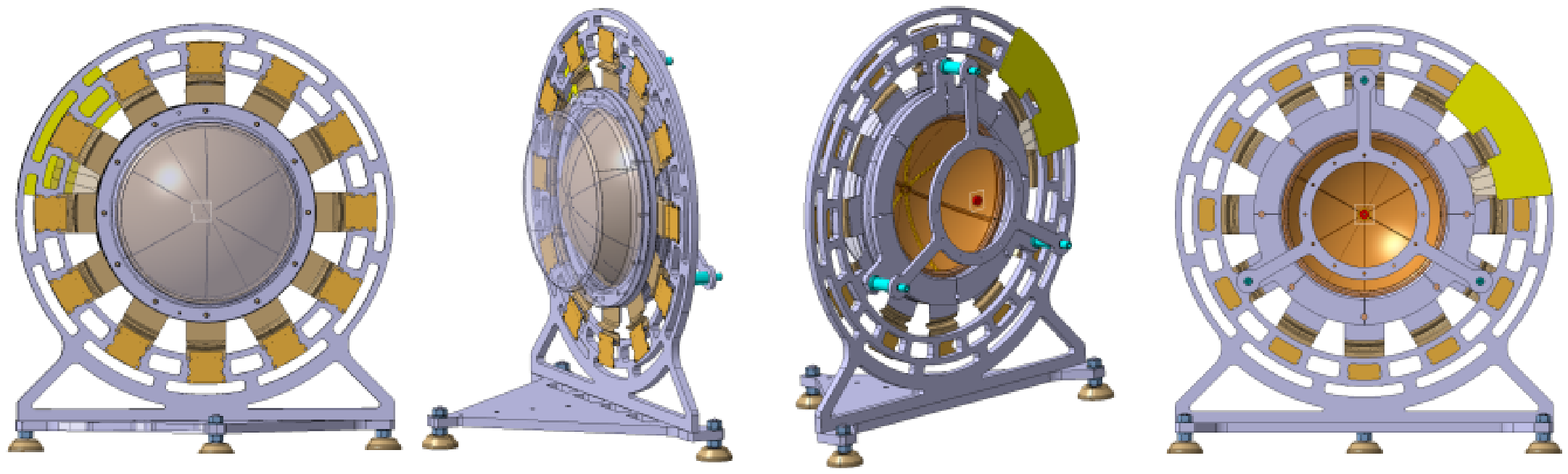}
\caption{3D design of a spherical triple \textsc{gem} detector from various points of view. The spherical active area has a diameter of $\sim 17$ cm, the whole structure is $\sim$ \oldstylenums{40} cm high.}
\label{3D}
\end{figure*}

The use of spacers in the drift region is considered, in case the spherical \textsc{gem} is not sufficiently self-supporting.
In case of a spherical multiple \textsc{gem} structure, spacers between the \textsc{gem}s cannot be avoided.
For planar \textsc{gem} detectors, spacers are easily made from plate materials.
The design is made with any \oldstylenums2\textsc{d} graphics package, and the plates are machined with a \textsc{cnc} bench.
For a spherical detector curved spacers are needed, which must be designed with \oldstylenums3\textsc{d} software.
Also the manufacturing is less straightforward.
The spacers for these prototypes are being made by \emph{stereolithography}, a fast \oldstylenums3\textsc{d} prototyping technique that uses a \textsc{uv} laser to polymerize a liquid epoxy in a selective and accurate way.

In order to arrive at a fully spherical detector, a spherical readout board must be developed.
This is a considerable technical challenge, as many possibilities are excluded: vias will crack, adhesives for multi-laminates do not support the high forming temperatures, standard rigid substrates cannot be formed like \textsc{gem}s.
We foresee a strip readout in a spiral pattern, where strips printed on top and bottom of a \textsc{gem}-like structure are mutually orthogonal (i.e. clockwise and counter-clock\-wise spirals), see Fig.~\oldstylenums{\ref{spirals}}.
The sharing of charge between top and bottom strips can be tuned by varying a low voltage between the sides.

\section{Conclusions and outlook}
\balance
\noindent We have shown that is it feasible to make spherical \textsc{gem}s, using a reasonably inexpensive method.
Many forming tests have been done to gain a good understanding of the parameters that lead to satisfying results, in a reproducible way.
The \textsc{gem}s made this way hold high voltage with the same low leakage as ordinary flat foils.
They will soon be tested with a spherical cathode and a planar readout structure.

The next challenge will be to develop a spherical readout board and make a fully spherical detector.
Detailed designs have already been prepared for this detector (see Fig.~\oldstylenums{\ref{3D}} on the next page), which will be the first entirely parallax free gas detector.

\end{document}